\documentclass[12pt,prl,aps,tighten,showpacs]{revtex4}
\usepackage{graphicx}
\usepackage{latexsym}
\usepackage{amsmath}
\usepackage{amssymb}
\usepackage{makeidx}

\begin{document}
\baselineskip=20pt
\title{The statistics of diffusive flux}

\author{Alba Margarita Res\'endiz Antonio$^{(1)}$, Hern\'an Larralde$^{(2)}$}

\affiliation{(1) Facultad de Ciencias, UAEMor. Av Universidad s/n, CP
  62210 Cuernavaca Morelos, Mexico\\
 (2) Centro de Ciencias F\'\i sicas, UNAM;
  Apartado postal 48-3, CP 62251 Cuernavaca Morelos, Mexico }
\begin{abstract}
We calculate the explicit probability distribution function for the
flux between sites a in a simple discrete-time diffusive system
composed by independent random walkers.  We highlight some of the
features of the distribution and we discuss its relation to the local
instantaneous entropy production in the system.  Our results are
applicable both to equilibrium and non equilibrium steady states as
well as for certain time dependent situations.
\end{abstract}
\maketitle 

Random walkers are used extensively as phenomenological models
for diffusive processes\cite{weiss}. The connection between physical
diffusion and random walks arises from the fact that the concentration
in a system containing many identical independent random walkers
satisfies Fick's law. That is, the mean flux between sites is
proportional to the difference in the mean number of walkers at each
site, plus, perhaps, a convective term if the walks are not
symmetric.  However, to the best of our knowledge, the statistics of
the flux in such a system have not been characterized (beyond the
mean, of course). This is not the case in more complicated systems, for
which a large amount of work on the statistics of currents has been
carried out.  Recent examples include statistics and large deviation
theory for the fluctuations of the current in lattice gases
\cite{bertini, bodineau, derrida}; the integrated current distribution
for the steady-state of the one-dimensional zero-range process
\cite{harris}; the joint probability function for the occupation
number and the current through the system in the asymmetric simple
exclusion process with open boundaries \cite{depken}, to mention but a
few.

In addition to being an alternative approach to the description of
transport properties of simple diffusive systems, the statistics of
flux are of relevance in many diffusion limited processes. For
example, in a diffusion reaction system in which the reactants are
initially separated, fluctuations in the flux of reactants into the
interfacial region give rise to fluctuations in the position of
the reaction front and in the net reaction kinetics
\cite{cox,sinder,araujo}.

Furthermore, viewed from the thermodynamic side, there has been
a great deal of work relating the statistics of flux to the entropy
production of non equilibrium steady states \cite{gallavotti,
kurchan,lebowitz}. From this perspective, diffusive systems modeled by
random walkers are among the simplest examples that can be
used to test and extend such ideas.

In this work we obtain an explicit expression for the probability
distribution of the flux, that is, the net number of particles that
hop between neighboring sites in a given time step, in a system
containing independent discrete time random walkers. This system has
the added bonus that the results are applicable to both equilibrium
and non-equilibrium situations, where the latter can be achieved by
imposing concentration differences on the boundaries or by subjecting
the system to an external field (by considering biased random walkers)
or both.

For definiteness, in this work we consider a discrete one dimensional
system in which independent discrete time random walkers evolve
synchronously according to the following rules: at each time step,
a walker moves to the site on its right or to the site on its left with
probabilities $p$ and $q$ respectively, or stays at its site with
probability $r=1-p-q$. We note that the discreteness in time allows us
to consider fluxes that arise from the simultaneous hopping of many
particles. Such multiparticle events are not present in continuous
time versions of the system; we discuss the continuous limit of the
problem further on.  

The system in which the random walkers evolve consists of $l+2$ sites
and we impose as boundary conditions that the number of walkers at
sites $i=0$ and $i=l+1$ be Poisson distributed with fixed mean values
$c_0$ and $c_{l+1}$ respectively \cite{footnote}.

To compute the probability distribution of particle flux between
neighboring sites, labeled $i$ and $i+1$, we require $J_{+}$ and
$J_{-}$, the number of particles that jump from site $i$ to site $i+1$
and the number of particles that jump from $i+1$ to $i$ respectively.
In terms of these, the total flux $J$ between these sites will be
given by $J=J_{+}-J_{-}$.

We denote by $m$ and $n$ the ocupancy of sites $i$ and $i+1$
respectively.  Then, the probability that the net flux between these
sites is $J$ can be expressed as
\begin{equation}
p_i(J)=\sum_{m,n}p(J|m,n)p_{i,i+1}(m,n),
\end{equation}
where $p_{i,i+1}(m,n)$ is the probability of finding exactly $m$ and
$n$ particles at sites $i$ and $i+1$, and $p(J|m,n)$ is the
probability of having a total flux $J$ between these sites given those
occupancy numbers.

Since the walkers are independent, we have
\begin{eqnarray}
p(J|m,n)&=&\sum_{J_{-}=0}^{\infty}p_{+}(J +J_{-}|m)p_{-}(J_{-}|n),\label{two}
\end{eqnarray}
where
\begin{equation}
 p_{+}(J_{+}|m)=
\begin{pmatrix} 
m \\
J_{+}
\end{pmatrix}p^{J_{+}}(1-p)^{m-J_{+}}
\end{equation}
and
\begin{equation}
p_{-}(J_{-}|n)=\begin{pmatrix}
  n \\
  J_{-}
\end{pmatrix}q^{J_{-}}(1-q)^{n-J_{-}}
\end{equation}
are the probabilities that $J_{+}$ particles jump to the right from a
site containing $m$ particles and  $J_{-}$ particles jump to the left from a
site containing $n$ particles.

In what follows, it will prove useful to work with the generating
function of expression (\ref{two}), defined as ${\hat p}(z|m,n)
:=\sum_{J=-\infty}^\infty z^{J} p(J|m,n)$, which is given
by
\begin{equation}
{\hat p}(z|m,n)=\left(1-q+\frac{q}{z}\right)^{n}(1-p+zp)^{m}.
\end{equation}

Next, we note that this system has the remarkable property that the
joint probability distribution $p_{i,i+1}(m,n)$ of the occupancy
numbers factorizes into a product of Poisson distributions
\cite{gaspard05}. This property is analogous to the factorization of
the steady state occupancy distribution which occurs in certain zero
range processes under similar boundary conditions \cite{schutz}. Thus,
the joint occupancy distribution $p_{i,i+1}(m,n)$ can be written as
\begin{eqnarray}
p_{i,i+1}(m,n)=\varphi_i(m)\varphi_{i+1}(n),
\end{eqnarray}
where
\begin{eqnarray}
\varphi_{i}(M)&=&\frac{1}{M!}e^{-c_{i}}c_{i}^{M},
\end{eqnarray}
and $c_{i}$ is the mean number of particles at site $i$. Actually,
under certain circumstances, namely initial conditions already
characterized by independent Poisson distributions, the joint
occupancy distribution can be expressed as the product of Poisson
distributions throughout the evolution of the system. In this case,
the mean occupation numbers can be obtained as the solution to the
discrete diffusion equation
\begin{equation}
c_i(t+1)-c_i(t)=pc_{i-1}(t)+qc_{i+1}(t)-(p+q)c_{i}(t)\label{conc}
\end{equation}
with appropriate initial and boundary conditions. However, just as for
zero range processes \cite{schutz}, the system always reaches a unique
factorizable steady state, independently of the initial conditions,
for which the parameters $c_i$ are given by
\begin{equation}
c_{i}=\frac{1}{\left(\frac{p}{q}\right)^{l+1}-1}
\left[c_{0}\left(\frac{p}{q}\right)^{l+1}-c_{l+1}+
(c_{l+1}-c_{0})\left(\frac{p}{q}\right)^{i} \right],
\label{steady}
\end{equation}
which is the steady state solution of equation (\ref{conc}).

\begin{figure}[h]
\begin{center}
\scalebox{0.8}{\includegraphics{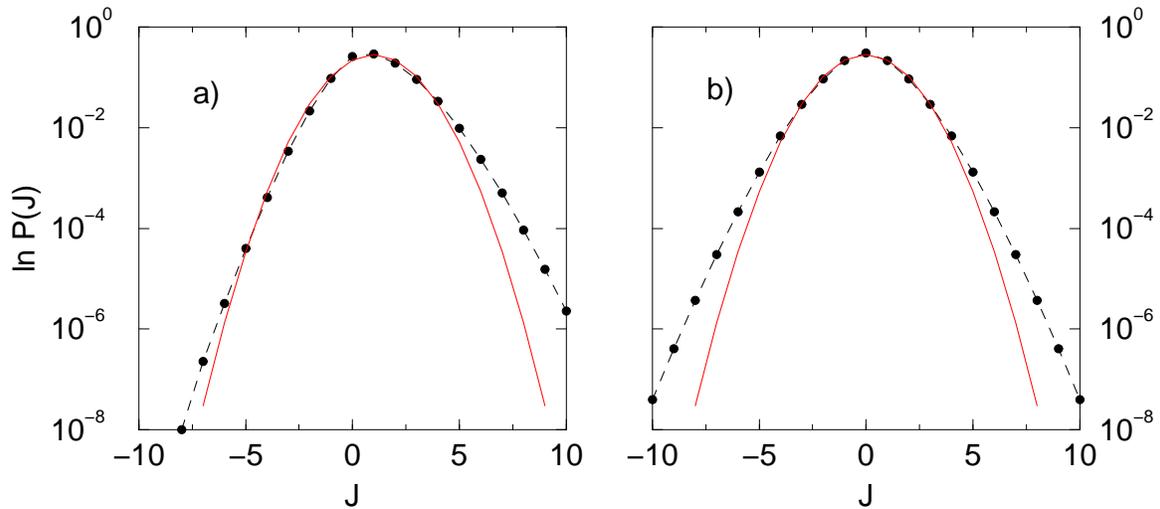}}
\end{center}
\caption{Semilogarithmic plots of $p_i(J)$ with variance equal to 2
and a) $\langle J\rangle=1$; b) $\langle J\rangle=0$ (dot-dashed
lines). Gaussian distributions with the same variance and mean as
$p_i(j)$ in each case are plotted for comparison.}
\label{fig}
\end{figure}

Thus, we are in a position to evaluate the generating function
$\hat p_i(z)$ of the distribution of fluxes between sites $i$ and $i+1$:
\begin{equation}
{\hat p_i(z)}=\sum_{m}\sum_{n}{\hat  p(z|m,n)}\varphi_i(m)\varphi_{i+1}(n)
=e^{-pc_{i}-qc_{i+1}+zpc_{i}+\frac{q}{z}c_{i+1}}.
\end{equation}
From this expression, the moments and the cumulants of the
distribution may be readily evaluated. For example, all the odd
cumulants are given by
\begin{equation}
\kappa_{2n-1}=pc_{i}-qc_{i+1},\qquad n=1,2,\ldots
\end{equation}
whereas the even cumulants are
\begin{equation}
\kappa_{2n}=pc_{i}+qc_{i+1},\qquad n=1,2,\ldots
\end{equation}
In particular, using the explicit expression for the concentration
profile for the steady states given in equation (\ref{steady}), we
find
\begin{equation}
\kappa_{2n-1}=\frac{1}{p^{l+1}-q^{l+1}}\left[(p-q)(c_{0}p^{l+1}-c_{l+1}q^{l+1})
\right]
\end{equation}
and
\begin{equation}
\kappa_{2n}=\frac{1}{p^{l+1}-q^{l+1}}\left[(p+q)(c_{0}p^{l+1}-c_{l+1}q^{l+1})
+2pq(c_{l+1}-c_{0})p^{i}q^{l-i}\right]
\end{equation}
Furthermore, using the explicit expressions for the moments we can
calculate the skewness $\gamma_1$ and excess (or kurtosis) $\gamma_2$
of the distribution, which are given by
\begin{equation}
\gamma_1=\frac{\langle( J-\langle
J\rangle)^{3}\rangle}{\sigma^{3}} 
=\frac{(pc_{i}-qc_{i+1})}{(pc_{i}+qc_{i+1})^{\frac{3}{2}}}
\end{equation}
and
\begin{equation}
\gamma_2=\frac{\langle( J-\langle
J\rangle)^{4}\rangle}{\sigma^{4}}-3
=\frac{1}{(pc_{i}+qc_{i+1})}.
\end{equation} 
Thus, only in the limit $pc_{i}+qc_{i+1}\gg 1$ is the distribution of
flux essentially Gaussian.  In fact, the expression for ${\hat
p_i(z)}$ can be inverted to obtain an explicit expression for
$p_i(J)$. Rewriting
\begin{equation}
{\hat p}_i(z)=e^{-pc_{i}-qc_{i+1}}
\exp\left[\frac{1}{2}(2\sqrt{pqc_{i}c_{i+1}})
\left(z\sqrt{\frac{pc_{i}}{qc_{i+1}}} +
\frac{1}{z\sqrt{\frac{pc_{i}}{qc_{i+1}}}}\right)\right],
\end{equation}
\noindent 
and comparing with the generating function of the modified Bessel
functions \cite{stegun}
\begin{equation}
\exp\left[{\frac{1}{2}}x\left(t+\frac{1}{t}\right)\right]
=\sum_{s=-\infty}^{\infty}I_{s}(x)t^{s},
\end{equation}
yields
\begin{equation}
p_i(J)=e^{-pc_{i}-qc_{i+1}} 
\left[\frac{pc_{i}}{qc_{i+1}}\right]^{\frac{J}{2}}I_{J} 
(2\sqrt{pqc_{i}c_{i+1}}).\label{dist}
\end{equation}
This distribution, shown in Fig.\ref{fig}, shares some of the
properties of the Gaussian distribution: It is characterized by only
two parameters, say $pc_{i}$ and $qc_{i+1}$ in this case. It satisfies
a generalized stability condition, in the sense that the distribution
of the sum of independent (integer) random variables distributed
according to $p_i(j)$, has the same form as $p_i(j)$ with appropriately
rescaled parameters. That this is the case is obvious from the
explicit expression of the generating function, but it could have been
foreseen from the actual process we are describing. Indeed, the
complete derivation presented above applies also for the distribution
of flux between adjacent cell in systems of arbitrary dimension. For
these, the one dimensional system we are considering corresponds to
a projection onto one of the axes. Thus, the flux distribution we
obtain can itself be thought of as that of the sum of many independent
fluxes with distributions similar to $p_i(j)$.  \vskip0.5cm
The large $|J|$ behavoir of $p_i(J)$, which is where it most
clearly differs from the Gaussian, is easily amenable to
evaluation. From the series expansion of the Bessel functions
\cite{stegun}, we immediately find that
\begin{eqnarray}
p_i(J)\sim\left\{
\begin{array}{ll} e^{-pc_{i}-qc_{i+1}}\frac{(pc_i)^J}{J!}\left(1+\frac{pqc_i
    c_{i+1}}{J} +\cdots\right) & ~~~~\mbox{when $J\to \infty$}\\
\\
e^{-pc_{i}-qc_{i+1}}\frac{(qc_{i+1})^{|J|}}{|J|!}\left(1+\frac{pqc_i
    c_{i+1}}{|J|} +\cdots\right) & ~~~~\mbox{when $J\to -\infty$.}\\
\end{array} \right.
\end{eqnarray}

Another limiting behavior of $p_i(J)$ worth mentioning is that
corresponding to the continuous time limit. This limit is achieved by
assuming that the hopping probabilities can be expressed as
$p=\alpha\delta t$ and $q=\beta\delta t$, where $\delta t$ is the time
interval between succesive steps. Then, keeping only the
terms up to linear order in $\delta t$ we find
\begin{eqnarray}
p_i(J)\approx\left\{
\begin{array}{ll} \alpha c_{i}\delta t & ~~~~\mbox{$J=1$}\\
1-(\alpha c_{i}+\beta c_{1+1})\delta t & ~~~~\mbox{$J=0$}\\
\beta c_{i+1}\delta t & ~~~~\mbox{$J=-1$.}\\
\end{array} \right.
\end{eqnarray}
All other values of $J$ have probabilities of higher order in $\delta
t$ and are, therefore, negligible as $\delta t \to 0$. The above
expression merely reflects the fact that in the continuum limit, the
transport of particles amongst neighboring sites during a time
interval $\delta t$ is, at most, a single particle process. The
statistical parameters in this limit can be calculated directly from
the expressions obtained above; thus, for example, the mean number of
particles that flow from site $i$ to site $i+1$ in a time interval
$\delta t$ is $(\alpha c_{i}-\beta c_{1+1})\delta t$ as was to be
expected.

A further characteristic that $p_i(J)$ shares with the Gaussian
distribution is that the ratio $p_i(J)/p_i(-J)$ is a pure
exponential. This quantity is of interest because it can be
associated to entropy production of non-equilibrium steady states.
Indeed, given the explicit form of $p_i(J)$ between sites $i$ and $i+1$
in a system, and the symmetry properties of the Bessel functions, we
have:
\begin{equation}
s_i=\ln\left(\frac{p_i(J)}{p_i(-J)}\right)=
J\ln\left(\frac{pc_i}{qc_{i+1}}\right)
\label{s}
\end{equation}
The average of this quantity is
\begin{equation}
\langle s_i\rangle = (pc_{i}-qc_{i+1})
\ln\left(\frac{pc_i}{qc_{i+1}}\right)\label{entropy}
\end{equation}
which has the familiar non-negative form that appears in the
{\it H}-theorem, and that we argue can be considered as an instantaneous
local entropy production for this system. It should be remarked that
this is not the same quantity that appears in the fluctuation theorems
\cite{gallavotti, kurchan, lebowitz}, which are concerned with the
large fluctuations of the time-integrated flux.

Also, from the definition of $s_i$ in equation (\ref{s}) it is again
apparent that this quantity is a random variable, the statistics of
which are simply related to the statistics of flux described
above. Thus, for example, the variance of $s_i$ will be given by
\begin{equation}
\langle s_i^2\rangle-\langle s_i\rangle^2=(pc_{i}+qc_{i+1})\left[
\ln\left(\frac{pc_i}{qc_{i+1}}\right)\right]^2
\end{equation}
To justify our identification of $\langle s_i\rangle$ as an average
local entropy production, first note that the equilibrium for this
system is defined as the steady state with zero average flux. In this
situation, the concentration profile will be given by
\begin{equation}
c_i=c_0 (p/q)^i,
\end{equation}
where $c_0$ is the concentration imposed on the boundary site $i=0$.
On the other hand, from the thermodynamic perspective the system
corresponds to noninteracting particles in an external field $h$. For
such a system the concentration profile is given by the barometric
equation
\begin{equation}
c_i=e^{\beta(\mu -ih)},
\end{equation}
where $\mu$ is the chemical potential and $\beta$ is the inverse
temperature. Comparing both expressions for the concentration profiles
at equilibrium leads to the identification
\begin{equation}
\beta\mu=\ln c_0;\qquad \beta h = \ln\frac{p}{q}.
\end{equation}
Of course, in general the system is not in equilibrium; however, since
the occupancy is characterized by independent Poisson distributions,
it can be considered as being in local equilibrium. Thus, we can
rewrite equation (\ref{entropy}) as:
\begin{equation}
\langle s_i\rangle=\langle J\rangle_i\left[\ln c_i - \ln c_{i+1}
  +\ln\frac{p}{q}\right]=-\langle
  J\rangle_i\left[\Delta_i\beta\left(\mu_i+\Phi(i)\right)
  \right],\label{prod}
\end{equation}
where $\Phi(i)=-ih$ is a ``potential energy'' and $\Delta_i$ is the
difference operator. Equation (\ref{prod}) is a discrete analog of the
local entropy production of linear thermodynamics. 

At this point it is worth emphasizing that the average entropy
production proposed in equation (\ref{entropy}) is obtained from a
{\it single} step local statistic. Thus, this quantity should be
interpreted as the entropy produced at that time step, averaged over
an ensemble of statistically identical systems, which is the usual
interpretation of statistical mechanics. This is especially relevant
in the time dependent case, for which all of the above results also
hold, as mentioned above, given that the initially the sites of each
system of the ensemble are occupied according to independent Poisson
distributions. For these it makes no sense to study time averaged
quantities of single systems and relate the results to the ensemble
averages. On the contrary, in the steady state the ensemble average of
$s_i$, Eq.~(\ref{s}), will coincide with the time average of the flux
in a single system, yielding the connection usually assumed by the
ergodic hypothesis. This will not be true for higher moments of $s_i$
due to correlations in the flux at different times.

If we restrict ourselves to the steady states, the average flux is
constant and we can calculate the total entropy production in the
system as
\begin{eqnarray}
s_{\rm total}&=&\sum\limits_{i=0}^l \langle s\rangle_i=\sum\limits_{i=0}^l
(pc_{i}-qc_{i+1})\ln\left(\frac{pc_i}{qc_{i+1}}\right)\\ \nonumber
&=&\frac{1}{p^{l+1}-q^{l+1}}\left[(p-q)(c_{0}p^{l+1}-c_{l+1}q^{l+1})\right]
\left[(l+1)\ln\left(p/q\right)+\ln\left(c_0/c_{l+1}\right)\right],
\end{eqnarray}
where $c_0$ and $c_{l+1}$ are the concentrations imposed on the
boundaries of the system and we used the explicit expression for the
steady state flux given in equation (\ref{steady}).  In particular,
if we take the limit $q=p$, the above expression coincides with that
obtained in \cite{gaspard05}, where the entropy production is
interpreted in terms of a time asymmetry in the dynamical randomness
between the forward and backward paths of the diffusion process
\cite{gaspard05,gaspard04}.

Finally, the limit in which $q\to0$ say, is worth discussing. This
totally asymmetric hopping case is again of some importance in the
context of zero range processes. The distribution of flux in this case
is easy to evaluate from the explicit expression in equation
(\ref{dist}), from which we obtain $p_i (J)=0$ for $J<0$ and
$p_i(J)$ is Poissonian with mean $pc_i$ for $J>0$.
Further, as discussed above, a field can be associated to
the $\ln(p/q)$, which gives rise to a ``potential energy''; thus,
in the limit when $q$ vanishes, motion against the field is impossible
and motion along the field ``dissipates'' an infinite amount of
energy. In accordance, the mean entropy production diverges in this
limit.

In summary, we have obtained the explicit probability distribution
function of single step flux between adjacent sites in a diffusive
system composed of independent discrete random walkers. The key to
solving the problem is the fact that the joint site occupancy
distribution for this system factorizes into independent Poisson
distributions even in non equilibrium situations. We also obtain the
statistical parameters that characterize the distribution and
discuss some of its limiting behaviors. In addition, we use the
distribution to evaluate a single step local entropy production, whose
average can be related to the usual expressions from nonequilibrium
thermodynamics. In this context, an interesting extension of this work
would be to introduce an additional conserved quantity which can be
distributed among the random walkers at each site. This energy-like
quantity may or may not affect the jumping rates (although from a
physical perspective it probably should), and would allow the study
of coupled transport phenomena in these extremely simple settings. We
have thus far been unable to pose such scenario in a tractable
manner. A different avenue of research would be to determine multiple
time and site flux statistics in systems containing many random
walkers, characterizing, for example, the correlations in flux at two
sites of the system as a function of time.

This work was partially supported by grant IN-100803 of DGAPA UNAM. We
thank D. Sanders,  F. Leyvraz and M. Aldana for the many useful comments and
suggestions on this manuscript.

\section{Appendix}
The key result that permits the explicit calculation of $p_i(J)$ is that
the joint occupation probability distribution of the system factorizes
into Poisson distributions. For the sake of completeness, we show how
this comes about. 

We require the determination of the joint occupation probability
distribution, that is, the probability
$p_t(n_{0},n_{1},\ldots,n_{l+1})$ of finding $n_0, n_1,\ldots,n_{l+1}$
particles, at the sites $0,1,\ldots,l+1$ respectively at time $t$,
with the boundary conditions described in the text. The
evolution equation for this quantity is
\begin{eqnarray}
p_{t+1}(n_{0},n_{1},\ldots,n_{l+1})&=&
e^{-c_{0}}\frac{c_{0}^{n_{0}}}{n_{0}!}
e^{-c_{l+1}}\frac{c_{l+1}^{n_{l+1}}}{n_{l+1}!}\left\lmoustache\right.
p_{t}(m_{0},m_{1},\ldots,m_{l+1})\\\nonumber
&\times&\prod_{i=1}^{l}
\delta(m+l_{i-1}^{+}+l_{i+1}^{-}-(l_{i}^{+}+l_{i}^{-})-n_{i})\\\nonumber
&\times&\prod_{i=0}^{l+1}
\begin{pmatrix}
m_{i}\\
l_{i}^{-}
\end{pmatrix}
\begin{pmatrix}
m_{i}-l_{i}^{-}\\l_{i}^{+}
\end{pmatrix}q^{l_{i}^{-}}p^{l_{i}^{+}}r^{m_{i}-l_{i}^{-}
-l_{i}^{+}},\label{long}
\end{eqnarray}
where
\noindent 
\begin{equation}\left\lmoustache\right.=
\sum_{m_{0}=0}^{\infty}\ldots\sum_{m_{l}=0}^{\infty}
\sum_{m_{l+1}=0}^{\infty}\sum_{l_{0}^{-}=0}^{\infty}
\ldots\sum_{l_{l}^{-}=0}^{\infty}\sum_{l_{l+1}^{-}=0}^{\infty}
\sum_{l_{0}^{+}=0}^{\infty}\sum_{l_{1}^{+}=0}^{\infty}\ldots
\sum_{l_{l+1}^{+}=0}^{\infty}\\
\nonumber
\end{equation}
and we recall that $r=1-p-q$. Equation (\ref{long}) looks messy, but
it is actually very easy to understand: if we consider the LHS of this
equation without the delta functions, we note that it contains all the
possible events from all the possible configurations at time $t$,
weighted by the probability that the events occur. The deltas restrict
this immense sum to the terms which end up in the occupancy
configuration specified on the RHS of the equation.

In expression (\ref{long}) we have already imposed Poisson occupancy
distributions with concentrations $c_0$ and $c_{i+1}$ on the sites $0$
and $l+1$ as boundary conditions for the system. This will enable us to
drive the system into ``nonequilibrium'' steady states by imposing
different concentrations on the boundary, or considering biased random
walks (i.e. $p\neq q$), or both.

This evolution equation can be dealt with by evaluating the
multivariate generating function of the joint distribution, namely:
\begin{equation}  
{\hat p}_{t}(z_{0},\ldots,z_{l+1})=\sum\limits_{n_0=0}^\infty
\sum\limits_{n_1=0}^\infty\ldots
\sum\limits_{n_{l+1}=0}^\infty z_0^{n_0} z_1^{n_1}\ldots z_{l+1}^{n_{l+1}}
p_{t}(n_{0},n_{1},\ldots,n_{l+1}).\label{gf}
\end{equation}
After some simple but tedious algebra, which consists in regrouping
terms and adding up binomial expansions, the generating function can
be shown to satisfy the recursion relation
\begin{eqnarray}  
{\hat p}_{t+1}(z_{0},\ldots,z_{l+1})&=&e^{-c_{0}(1-z_{0})} 
e^{-c_{l+1}(1-z_{l+1})}
 \\\nonumber
&\times&\sum_{m_{0}=0}^{\infty}\ldots\sum_{m_{l+1}=0}^{\infty}p_{t}(m_{0}
,m_{1},\ldots,m_{l+1})\\\nonumber
&\times&[1+(z_{1}-1)p]^{m_{0}}[q+z_{2}p+z_{1}r]^{m_{1}}\\\nonumber
&\times&[z_{l-1}q+p+z_{l}r]^{m_{l}}[1+(z_{l}-1)q]^{m_{l+1}}\\
\nonumber
&\times&\prod_{i=2}^{l-1}[z_{i+1}p+z_{i-1}q+z_{i}r]^{m_{i}}.        
\end{eqnarray}
Comparing with the definition of the generating function, we note that
this expression is equivalent to
\begin{eqnarray}\label{occ-rec}
  {\hat p}_{t+1}(z_{0},z_{1},\ldots,z_{l},z_{l+1})=e^{-c_
{0}(1-z_{0})}e^{-c_{l+1}(1-z_{l+1})}&\times&\\\nonumber {\hat
p}_{t}\big(1+(z_{1}-1)p,~z_{2}p+q+z_{1}r,~z_{3}p\!&+&\!\!z_{1}q+z_{2}r,
\ldots\\\nonumber
\ldots,~z_{l}p+z_{l-2}q+z_{l-1}r,~z_{l-1}q\!&+&\!\!p+z_{l}r,~1+(z_{l}-1)q\big).
\end{eqnarray}                                         

To solve this equation we make the ansatz that the
distribution is given by the product of Poisson distributions, and
thus, its generating function can be written as
\begin{equation}
{\hat p}_{t}(z_{0},\dots,z_{l+1})=e^{-\sum_{i=0}^{l+1}c_{i}(t)
(1-z_{i})}.
\end{equation}
\noindent 
Substitution of this expression into Eq.(\ref{occ-rec}) leads to the
following equations for the mean occupation numbers $c_i(t)$ which
characterize the Poisson distributions at each site:
\begin{eqnarray}
\sum\limits_{i=0}^{l+1}&&\!\!\!\!\!\!\!\!\!c_{i}(t+1)[1-z_{i}]=c_{0}[1-z_{0}]+
c_{l+1}[1-z_{l+1}]
+ c_{0}(t)p[1-z_{1}]+ c_{1}(t)[1-(z_{2}p+q+z_{1}r)]\nonumber\\
&+&\sum\limits_{i=2}^{l-1}c_{i}(t)[1-(z_{i+1}p+z_{i-1}q+z_{i}r)]
+c_{l}(t)[1-(z_{l-1}q+p+z_{l}r)]
+c_{l+1}(t)q[1-z_{l}]\nonumber
\end{eqnarray}
\noindent 
If we now impose the boundary conditions
$c_{0}(t)=c_{0}$ and $c_{l+1}(t)=c_{l+1}$ for all times $t$, the above
equation may be rewritten as
\begin{eqnarray}                                     
&&\sum_{i=1}^{l}c_{i}(t+1)-\sum_{i=1}^{l}c_{i}(t) =
(c_{0}p-c_{1}(t)q)-(c_{l}(t)p-c_{l+1}q)\\\nonumber &+&
\sum_{i=1}^{l}[c_{i}(t+1)-c_{i-1}(t)p-c_{i+1}(t)q-c_{i}(t)+c_{i}(t)(p+q)]z_{i}
\end{eqnarray}
\noindent 
which must be valid for all values of the variables $z_{i}$. This is
achieved if, in addition to the boundary conditions, the parameters
$c_i(t)$ satisfy the discrete diffusion equation
\begin{eqnarray}
c_{i}(t+1)-c_{i}(t)&=&pc_{i-1}(t)+qc_{i+1}(t)-(p+q)c_{i}(t).\label{conc-diff}
\end{eqnarray}
Thus, if we have an ensemble of systems in which each
site is initially occupied according to independent Poisson
distributions, then the occupancy distribution will remain being the
product of Poisson distributions throughout the evolution of the
process.

Furthermore, independently of the initial conditions, the system will
reach a unique stationary state characterized by the product
distribution with the parameters $c_{i}$ solutions of the
stationary state of equation (\ref{conc-diff}):

\begin{equation}
c_{i}=\frac{1}{\left(\frac{p}{q}\right)^{l+1}-1}
\left[c_{0}\left(\frac{p}{q}\right)^{l+1}-c_{l+1}+
(c_{l+1}-c_{0})\left(\frac{p}{q}\right)^{i} \right].
\end{equation}

With this expression, we can calculate the explicit values of the
mean flux and its variance as functions of $p$, $q$, the imposed
boundary concentrations $c_0$ and $c_{l+1}$, and, for the variance, the
position along the system:
\begin{equation}
\langle J_{N}\rangle =
\frac{1}{p^{l+1}-q^{l+1}}\left[(p-q)(c_{0}p^{l+1}-c_{l+1}q^{l+1}) \right] 
\end{equation}

\begin{equation}
\sigma_{i}^{2}=\frac{(p+q)}{(p-q)}\langle
J_{N}\rangle+2pq\left(\frac{c_{l+1}-c_{0}}{p^{l+1}-q^{l+1}}\right)p^{i}q^{l-i}.
\end{equation}

Finally, it is worth mentioning that the above derivation can also be
extended to include site dependent jump rates.


\begin{thebibliography}{99}
\bibitem{footnote} This is a natural choice of boundary conditions if
  we wish to model contacts with a chemical baths at equilibrium, for
  which the number of particles is Poisson distributed in the grand
  canonical ensemble. Further, as shown in the appendix, this choice
  allows the exact factorizability of the joint occupancy
  distribution.
\end{thebibliography}
\end{document}